\documentclass[conference]{IEEEtran}
\IEEEoverridecommandlockouts

\usepackage{balance}
\usepackage{booktabs}
\usepackage{cite}
\usepackage{float}
\usepackage[acronym]{glossaries}
\usepackage{graphicx}
\PassOptionsToPackage{hyphens}{url}
\usepackage[colorlinks, citecolor=black, filecolor=black, linkcolor=black, urlcolor=black,
  pdftitle={Coding of volumetric content with MIV using VVC subpictures},
  pdfauthor={M. Santamaria, V. K. Malamal Vadakital, L. Kondrad, A. Hallapuro, M. M. Hannuksela}]{hyperref}
\usepackage{makecell}
\usepackage{multirow}
\usepackage[caption=false]{subfig}
\usepackage{tikz}
\usetikzlibrary{fit, positioning}
\usepackage{url}

\setacronymstyle{long-short}

\newacronym{3d}{3D}{three-dimensional}
\newacronym{3dof}{3DoF}{three degrees of freedom}
\newacronym{6dof}{6DoF}{six degrees of freedom}
\newacronym{acr}{ACR}{Adaptive Resolution Change}
\newacronym{bd-rate}{BD-rate}{Bj{\o}ntegaard Delta rate}
\newacronym{cg}{CG}{Computer Generated}
\newacronym{cmp}{CMP}{Cube Map Projection}
\newacronym{ctc}{CTC}{Common Test Conditions}
\newacronym{ctu}{CTU}{Coding Tree Unit}
\newacronym{cu}{CU}{Coding Unit}
\newacronym{erp}{ERP}{Equirectangular Projection}
\newacronym{fov}{FoV}{Field-of-View}
\newacronym{hd}{HD}{High Definition}
\newacronym{hevc}{HEVC}{High Efficiency Video Coding}
\newacronym{iec}{IEC}{International Electrotechnical Commission}
\newacronym{irap}{IRAP}{Intra Random Access Point}
\newacronym{iso}{ISO}{International Organization for Standardization}
\newacronym{itu-t}{ITU-T}{International Telecommunication Union - Telecommunication}
\newacronym{iv-psnr}{IV-PSNR}{Immersive Video PSNR}
\newacronym{mbps}{Mbps}{Megabits per second}
\newacronym{mcts}{MCTS}{Motion Constrained Tile Set}
\newacronym{miv}{MIV}{MPEG Immersive Video}
\newacronym{mpeg}{MPEG}{Motion Picture Experts Group}
\newacronym{nal}{NAL}{Network Abstraction Layer}
\newacronym{nc}{NC}{Natural Content}
\newacronym{omaf}{OMAF}{Omnidirectional Media Format}
\newacronym{psnr}{PSNR}{Peak Signal-to-Noise Ratio}
\newacronym{qp}{QP}{Quantisation Parameter}
\newacronym{sei}{SEI}{Supplemental Enhancement Information}
\newacronym{tmiv}{TMIV}{Test Model of Immersive Video}
\newacronym{uhd}{UHD}{Ultra-High Definition}
\newacronym{v3c}{V3C}{Visual Volumetric Video-based Coding}
\newacronym{vceg}{VCEG}{Video Coding Experts Group}
\newacronym{vcl}{VCL}{Video Coding Layer}
\newacronym{vmaf}{VMAF}{Video Multimethod Assessment Fusion}
\newacronym{vps}{VPS}{V3C Parameter Set}
\newacronym{vtm}{VTM}{VVC Test Model}
\newacronym{vvc}{VVC}{Versatile Video Coding}
\newacronym{ws-psnr}{WS-PSNR}{Weighted-to-Spherically-uniform PSNR}

\def\BibTeX{{\rm B\kern-.05em{\sc i\kern-.025em b}\kern-.08em
  T\kern-.1667em\lower.7ex\hbox{E}\kern-.125emX}}

\begin{document}

\title{Coding of volumetric content with \gls{miv} using \gls{vvc} subpictures}

\author{
  \IEEEauthorblockN{
    Maria Santamaria\IEEEauthorrefmark{1},
    Vinod Kumar Malamal Vadakital\IEEEauthorrefmark{2},
    Lukasz Kondrad\IEEEauthorrefmark{2}, \\
    Antti Hallapuro\IEEEauthorrefmark{2} and
    Miska M. Hannuksela\IEEEauthorrefmark{2}}
  \IEEEauthorblockA{\IEEEauthorrefmark{1}Tampere University, Tampere, Finland}
  \IEEEauthorblockA{\IEEEauthorrefmark{2}Nokia Technologies, Tampere, Finland}
}

\maketitle

\IEEEpubidadjcol

\begin{abstract}
Storage and transport of \gls{6dof} dynamic volumetric visual content for
immersive applications requires efficient compression. \acrshort{iso}/\acrshort{iec}
\acrshort{mpeg} has recently been working on a standard that aims to efficiently
code and deliver \gls{6dof} immersive visual experiences. This standard is called
the \gls{miv}. \Gls{miv} uses regular 2D video codecs to code the visual data.
\Acrshort{mpeg} jointly with \acrshort{itu-t} \acrshort{vceg}, has also
specified the \gls{vvc} standard. \Gls{vvc} introduced recently the concept of
subpicture. This tool was specifically designed to provide independent
accessibility and decodability of sub-bitstreams for omnidirectional
applications. This paper shows the benefit of using subpictures in the \gls{miv}
use-case. While different ways in which subpictures could be used in \gls{miv}
are discussed, a particular case study is selected. Namely, subpictures are used
for parallel encoding and to reduce the number of decoder instances. Experimental
results show that the cost of using subpictures in terms of bitrate overhead is
negligible (0.1\% to 0.4\%), when compared to the overall bitrate. The number
of decoder instances on the other hand decreases by a factor of two.
{\let\thefootnote\relax\footnote{This document is the accepted version of the paper that has been published as:
M. Santamaria, V. K. Malamal Vadakital, L. Kondrad, A. Hallapuro and M. M. Hannuksela,
"Coding of volumetric content with MIV using VVC subpictures," 2021 IEEE 23rd
International Workshop on Multimedia Signal Processing (MMSP), 2021, pp. 1-6, doi:
10.1109/MMSP53017.2021.9733465. Online:
\url{https://ieeexplore.ieee.org/abstract/document/9733465} \\
\copyright 2021 IEEE. Personal use of this material is permitted. Permission from IEEE must be obtained
for all other uses, in any current or future media, including reprinting/republishing this material for
advertising or promotional purposes, creating new collective works, for resale or redistribution to
servers or lists, or reuse of any copyrighted component of this work in other works.
}}
\end{abstract}


\glsresetall

\section{Introduction}
\label{sec:intro}
An immersive \gls{6dof} representation, unlike a \gls{3dof} representation,
provides a larger viewing-space, where viewers have both translational and
rotational freedom of movement at their disposition. In a \gls{3dof} visual
experience, content is presented to viewers as if they were positioned at the
centre of a sphere, looking outwards, with all parts of the content positioned
at some constant depth. Contrarily to \gls{3dof}, \gls{6dof} videos enable
perception of motion parallax, where the relative positions of scene geometry
change with the pose of the viewer. The absence of motion parallax in \gls{3dof}
videos is inconsistent with the workings of a normal human visual system and
often leads to visual discomfort~\cite{cit:Blinder2019}.

The large dimensions of a \gls{6dof} viewing-space increases the amount of data
required to describe the volumetric scene. Hence, the \gls{iso}/\gls{iec}
\gls{mpeg} is specifying the \gls{miv} standard~\cite{cit:miv} to efficiently
code dynamic volumetric visual scenes. This standard caters to virtual reality,
augmented reality and mixed reality applications, such as gaming, sports
broadcasting, motion picture productions, and telepresence.

\Gls{miv} defines the syntax and semantics of information that enable rendering
a viewport of a \gls{3d} scene from associated video coded data. The standard
supports scenes where the viewing-space is slightly larger than the scale of
motion of a human head~\cite{cit:Jung2020}. A viewing-space, in this context,
is a \gls{3d} sub-space of the captured scene, from within which a \gls{6dof}
experience can be rendered without significant artifacts.

\Gls{miv} can also accommodate volumetric scenes captured by a wide variety of
multi-camera arrangements, e.g., spherical, dome, and arrays. \gls{erp},
perspective, and orthogonal projection formats are supported. The \gls{miv}
bitstream consists of video coded data and non-video coded information related
to the coded video data. Non-video coded information related to coded video
data includes the projection format, camera parameters, depth \glspl{qp}, and
details of patches packed in video frames. The encoding of video data is left
to regular video codecs, such as the \gls{hevc}~\cite{cit:hevc} or the
\gls{vvc}~\cite{cit:vvc}.

The \gls{vvc} standard, a joint effort by \gls{itu-t} \gls{vceg} and \gls{iso}
\gls{mpeg}, was finalised in July 2020. This standard has the designation H.266
in \gls{itu-t} and \gls{iso}/\gls{iec} 23090-3 in \gls{iso}/\gls{iec}. \gls{vvc}
is a block-based hybrid video coding scheme that can achieve \( 50\% \) bitrate
reduction compared to \gls{hevc}~\cite{cit:Bross2021}. Apart from an increased
coding efficiency, \gls{vvc} also introduces new and versatile tools useful in
a variety of applications, e.g., coding of \gls{uhd} video content,
high dynamic range content, screen content, and omnidirectional content.

Subpicture is a new picture partitioning scheme included in \gls{vvc}. A subpicture
is a coded rectangular region of a picture, which is either extractable
(coded independently) or non-extractable. The former can be extracted using
a sub-bitstream extraction process, and can be merged with other subpicture
sub-bitstreams~\cite{cit:Wang2021}. Functionally, subpictures are similar to
\gls{mcts}~\cite{cit:Hannuksela2004} in \gls{hevc}, as both allow the extraction
of parts of a coded picture. Subpicture is a useful tool for \gls{3dof} and
\gls{6dof} use-cases, where only a part of a complete scene is rendered at a
given point in time.

This paper shows the benefit of utilizing subpicture in \gls{6dof} context,
particularly for coding \gls{miv} video data. Simulation results show that
using the independently composable and extractable property of subpicture has
negligible impact on bitrate of \gls{miv} bitstream.

The rest of the paper is organised as follows. Section~\ref{sec:sota} presents
the related work. Sections~\ref{sec:miv} and~\ref{sec:vvc} describe the \gls{miv}
standard and the \gls{vvc} standard, respectively. Section~\ref{sec:miv-subpic}
summarises how subpicture can be used in \gls{vvc}. Section~\ref{sec:approach}
introduces the packed representation of the video coded data.
Section~\ref{sec:experiments} shows the experimental results. Finally,
Section~\ref{sec:conclusion} contains some final remarks.

\section{Related work}
\label{sec:sota}
\Gls{vvc} has been used to reduce the overall bitrate in viewport-adaptive
streaming of omnidirectional video. For instance, Homayouni et
al.~\cite{cit:Homayouni2020} took advantage of the \gls{vvc} capability of
having mixed types of subpictures within a coded picture to cope with user's
head motion. In this approach, each subpicture sequence is encoded using long
and short \gls{irap} periods. The long \gls{irap} period is used on subpictures
where the viewport change does not affect the quality, whilst the short
\gls{irap} period is used on the remaining subpictures. The capability of mixing
long and short \gls{irap} periods in the same received bitstream reduces the
number of intra-coded areas in the received bitstream and thereby decreases
the bitrate. Adaptive video streaming is also tackled by Carreira et
al.~\cite{cit:Carreira2020}, but using the \gls{acr} concept from \gls{vvc}.
\Gls{acr} is used to encode a \gls{fov} as multiple frames with different
spatial resolution. The high resolution is destined for salient \gls{fov},
whereas low resolution is utilised for less relevant content. Another work
from Carreira et al.~\cite{cit:Carreira2019} maps \gls{cmp} to frame using the
temporal scalability of \gls{vvc}. Each cube face is a \gls{fov} and a frame
that is encoded in a different temporal layer than other \glspl{fov}. Moreover,
the defined coding structure disables the dependencies between different
\glspl{fov}. On the other hand, Skupin et al.~\cite{cit:Skupin2020} introduced
a rate control technique that operates at tile level. The goal in this case is
to have a fair quality distribution across all tiles, for which a random forest
model is used to reduce the \gls{qp} variability.

Adhuran et al.~\cite{cit:Adhuran2020} proposed a spherical adaptive objective
function that aims to reduce redundant data in \glspl{erp} while reducing the
quality loss. The objective function is minimised to derive the optimal
\gls{qp}, which along the function itself are used in the actual encoding process.

Other works focused on parallel processing. For example, Filipe et
al.~\cite{cit:Filipe2020} proposed a slice-based splitting algorithm to
balance the processing load across multiple processors. The algorithm calculates
the computational complexity of a block. The estimations are used to split the
\gls{3dof} video into slices with different sizes but similar computational
complexity.

The previous approaches make use of \gls{vvc} to code \gls{3dof} content. This
paper focuses on how to make use of \gls{vvc} and \gls{miv} to code \gls{6dof}
content.

\section{The MPEG Immersive Video standard}
\label{sec:miv}
The inputs to an \gls{miv} encoder are multiple sets of videos, captured by an
unordered group of real or virtual cameras having an arbitrary pose
(source-views). The set of videos from each source-view represents a projection
of a part of a volumetric scene onto the camera projection plane. Each video
referenced from a source-view describes either projected geometry (depth) or
attributes, e.g., texture, normal, and material map. The geometry video can be
generated using dedicated depth sensors or computed using computer vision
techniques. The technology used to capture/estimate geometry data is outside
the scope of the standard.

Schematically, the \gls{miv} encoding process starts with an analysis of the
video sets from source-views and their corresponding camera parameters. The
analysis step involves the partitioning of the set of source-views into a set
of basic-views and a set of additional-views. A basic-view, in this context, is
a source-view left unmodified. Additional-views, on the other hand, are
candidates considered for the pruning step that follows.

The multi-view inputs to \gls{miv} contain significant pixel level redundancies,
which the pruning step exploits. In the pruning step, each pixel of the
basic-views is first un-projected into the world space, and then reprojected
back onto the additional-views. A pixel of an additional-view is marked to be
pruned, if it is deemed to be redundant, i.e., the values of geometry and
texture are very close to the values in the basic-view. The order in which
additional-views are pruned is stipulated by using a hierarchical structure
called the pruning graph. To keep complexity manageable, the pruning graph is
generated using a greedy algorithm. The process of marking a pixel to be pruned
or otherwise generates a binary mask. To preserve temporal consistency, the mask
is aggregated over some number of frames. The aggregated masks then go through a
process of clustering to obtain coherent spatial regions (without holes).
Patches are generated from this pruning process and the resulting patches are
finally packed into one or more video frames, refer to as atlases.

Atlases are compressed with a video codec, such as \gls{vvc}, and stored as
\gls{v3c}~\cite{cit:v3c} video components. The metadata describing the patches
in atlases is signalled using the \gls{v3c} atlas component syntax, which
provides functionalities and similar flexibility as available in many video
coding specifications. For instance, it provides sequence and frame parameter
sets, it may be divided into tiles, and it is encapsulated into \gls{nal} units.

In order to correctly associate \gls{v3c} atlas components (i.e., encoded
metadata) and \gls{v3c} video components (i.e., encoded atlases), \gls{miv}
stores all \gls{v3c} components as a \gls{v3c} sequence, which is a sequence of
\gls{v3c} units. A \gls{v3c} unit header contains an indication of the type of
data (i.e., component) to follow in \gls{v3c} unit payload. Each \gls{v3c} unit
carries data belonging to one component (atlas or video). The first \gls{v3c}
unit in a \gls{v3c} sequence is a \gls{vps} that provides information about
profile and level of the sequence as well mapping of video components to codecs.

The \gls{vps} also provides signalling information that allows mapping a number
of \gls{v3c} video components into one \gls{v3c} video component, which is
called the \gls{v3c} packed video component.

\section{The Versatile Video Coding standard}
\label{sec:vvc}
\gls{vvc} uses a block-based hybrid video coding design, meaning the unit of
processing is a block of pixels rather than the whole frame. Moreover, the
coding combines predictive coding and transform coding of the prediction error.

A \gls{vvc} bitstream consists of a series of \gls{nal} units, which can be
\gls{vcl} or non-\gls{vcl} \gls{nal} units. \gls{vcl} \gls{nal} units contain
values of colour component samples. Coded data within the \gls{vcl} \gls{nal}
units is known as slices. Non-\gls{vcl} \gls{nal} units include data different
than slices, like parameter set \gls{nal} units and \gls{sei} \gls{nal} units.

A frame is first split into \glspl{ctu}, where a \gls{ctu} is a squared region
of luma and chroma samples. Next, each \gls{ctu} is split into \glspl{cu} using
a multi-type tree that considers quad-tree, binary, and tertiary trees. Each
\gls{cu} is further split until the minimum supported size is reached. A
\gls{cu} is the basic processing unit for operations such as predictive
coding and transform coding. Predictive coding removes redundancies within a
frame (spatial) and between frames (temporal). Transform coding decorrelates
the prediction error so less important coefficients can be discarded.

The high-level partitioning of a frame is composed of slices, tiles, and
subpictures, each comprising a number of complete \glspl{ctu}. A slice contains
coded content that can be reconstructed independently from other slices within
the same frame, mostly since in-frame prediction and entropy coding dependencies
are disabled across the slice boundaries. The same applies to tiles, which
specify horizontal and vertical boundaries and split a frame into columns and
rows. Additionally, each tile can be processed by a single processor core.
Finally, subpicture is a rectangular set of slices that can be decoded
independently of other subpictures (marked as extractable).

Subpictures have the following properties. First, the encoder can determine
the subpicture boundaries to be treated like picture boundaries in inter
prediction. Consequently, the samples at subpicture boundaries are padded when
motion vectors reference sample locations outside subpicture boundaries. Second,
independent subpictures can be extracted from their source bitstream and merged
into a destination bitstream without the need for rewriting slice headers.
Third, subpictures of the same coded picture need not have the same \gls{vcl}
\gls{nal} unit type. This enables extracting a first set of subpictures from
an \gls{irap} picture and a second set of subpictures from a non-\gls{irap}
picture and merging both sets into the same coded picture in a destination
bitstream.

\section{Ways to use subpictures in MIV}
\label{sec:miv-subpic}
Subpicture allows the creation of efficient bitstream with independently
extractable parts. The signalling overhead to indicate the independent
accessibility is marginal, considering the large amounts of data that is to be
coded. The use of subpicture allows: (1) viewport-dependent decoding and
rendering; (2) scalable rendering of either a \gls{3dof} or \gls{6dof} variation
of the content from a single bitstream; (3) parallel encoding of atlases; and
(4) minimising the number of streams to be synchronously decoded and rendered.

A viewport of an omnidirectional picture/sequence is the rectilinear projection
of the contents in the scene that is presented to a viewer. The portion of a
scene that is outside the viewport is not rendered for display. In an ideal
case, only the parts of the bitstream that are in the viewport require decoding
and rendering. However, practically, at least a low quality version of the
bitstream is also made available to the client, to account for unpredictable
viewer head motion. Since, minimally, only the viewport is required to be
decoded and rendered at high quality, subpictures can be used to partition the
omnidirectional visual field. For example, an omnidirectional scene can be
projected using \gls{cmp}, and each cube face can be coded as a subpicture.
Consequently, only the subpictures that fall in the viewport are required to be
decoded and rendered, reducing the computational resources needed.

Immersive video is still an active field of research and commercial product
development. Presently, a majority of display devices can only render \gls{3dof}
video. The \gls{miv} specification is flexible to allow generating bitstreams to
cater to both \gls{3dof} as well as a \gls{6dof} capable devices. An atlas can
be created such that one subpicture of the atlas represents a full \gls{erp}
image/video captured by one of the cameras, and the remaining subpictures
represents the additional depth information and patches required for \gls{6dof}
viewing.

\begin{figure}[t]
  \centering
  \includegraphics[width=.48\textwidth]{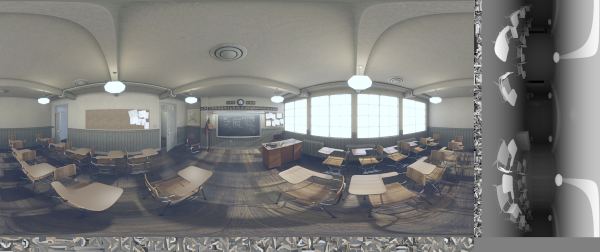}
  \caption{Packed texture and geometry.}
  \label{fig:packed-atlases}
\end{figure}

Fig.~\ref{fig:packed-atlases} illustrates a frame of an atlas that is
created in this manner. The independently extractable property of a subpicture
would allow either a network element, e.g. an edge computing server, or a even
a client device to choose the subpicture that is required for proper rendering
of a viewport. Players that only support \gls{3dof} video can identify the
\gls{erp} texture part of the decoded pictures using the region-wise packing
information that can be present in the video bitstream as a supplemental
enhancement information message and/or in the file metadata specified in the
\gls{omaf}~\cite{cit:Hannuksela2021}. Consequently, upgrading immersive video
services with \gls{miv} and \gls{6dof} rendering can be done without comprising
the compatibility with legacy players and devices only supporting \gls{3dof}
\( 360^\circ \) video.

A video encoder is complex, and encoding a frame of video can take a non-trivial
duration of time. Parallelism in video coding is not a new concept and there are
several tools that have been specified (wavefront parallel processing, tiles,
etc.). Subpicture is another tool to enable parallelism. For example, in the
\gls{miv} case, the geometry atlases and the texture atlases can be encoded
independently.

In practice, synchronised decoding and rendering of multiple coded video
bitstream is challenging. If even a single frame of the video bitstream is not
decoded and available in time, the rendered output of the volumetric video is
going to be incorrect. This problem can be handled by making the decoded frames
available in advance, which would require client-side buffering. However, memory
is a limited resource, especially in low-power mobile devices. Therefore,
reducing the number of coded bitstreams to be decoded would help with this
synchronisation problem. In section VI, it is shown that by packing an atlas
with both the texture and geometry components, a single decoder can be used,
thus solving the synchronisation problem.

\section{Packed representation of an atlas}
\label{sec:approach}
The goal is to reduce the number of \gls{vvc} decoder instances required to
reconstruct the compressed \gls{miv} atlases. For that purpose, texture and
geometry atlases are encoded with constraints that allow to combine them into
one bitstream, where each of them represents a single sub-bitstream.

\begin{figure*}[t!]
  \centering
  \resizebox{.75\textwidth}{!}{%
  \tikzset{every picture/.style={line width=0.75pt}} %set default line width to 0.75pt

\begin{tikzpicture}[x=0.75pt,y=0.75pt,yscale=-1,xscale=1]
%uncomment if require: \path (0,551); %set diagram left start at 0, and has height of 551

%Straight Lines [id:da14784817146046103]
\draw    (54.5,37.18) -- (65,37.18) -- (77,37.18) ;
\draw [shift={(80,37.18)}, rotate = 180] [fill={rgb, 255:red, 0; green, 0; blue, 0 }  ][line width=0.08]  [draw opacity=0] (8.93,-4.29) -- (0,0) -- (8.93,4.29) -- cycle    ;
%Straight Lines [id:da1147291795601676]
\draw    (231.66,53.68) -- (254.5,53.68) ;
\draw [shift={(257.5,53.68)}, rotate = 180] [fill={rgb, 255:red, 0; green, 0; blue, 0 }  ][line width=0.08]  [draw opacity=0] (8.93,-4.29) -- (0,0) -- (8.93,4.29) -- cycle    ;
%Straight Lines [id:da4463406506702604]
\draw    (410.41,53.68) -- (434,53.68) ;
\draw [shift={(437,53.68)}, rotate = 180] [fill={rgb, 255:red, 0; green, 0; blue, 0 }  ][line width=0.08]  [draw opacity=0] (8.93,-4.29) -- (0,0) -- (8.93,4.29) -- cycle    ;
%Straight Lines [id:da9792824647551537]
\draw    (333.5,53.68) -- (357,53.68) ;
\draw [shift={(360,53.68)}, rotate = 180] [fill={rgb, 255:red, 0; green, 0; blue, 0 }  ][line width=0.08]  [draw opacity=0] (8.93,-4.29) -- (0,0) -- (8.93,4.29) -- cycle    ;
%Straight Lines [id:da7895296771054308]
\draw    (54.5,66.35) -- (65,66.35) -- (77,66.35) ;
\draw [shift={(80,66.35)}, rotate = 180] [fill={rgb, 255:red, 0; green, 0; blue, 0 }  ][line width=0.08]  [draw opacity=0] (8.93,-4.29) -- (0,0) -- (8.93,4.29) -- cycle    ;
%Straight Lines [id:da6515118387214495]
\draw    (155.39,37.09) -- (166.63,37.09) -- (178.3,37.09) ;
\draw [shift={(181.3,37.09)}, rotate = 180] [fill={rgb, 255:red, 0; green, 0; blue, 0 }  ][line width=0.08]  [draw opacity=0] (8.93,-4.29) -- (0,0) -- (8.93,4.29) -- cycle    ;
%Straight Lines [id:da7048330699750759]
\draw    (54.5,122) -- (65,122) -- (77,122) ;
\draw [shift={(80,122)}, rotate = 180] [fill={rgb, 255:red, 0; green, 0; blue, 0 }  ][line width=0.08]  [draw opacity=0] (8.93,-4.29) -- (0,0) -- (8.93,4.29) -- cycle    ;
%Straight Lines [id:da33869080628567905]
\draw    (231.66,139.24) -- (254.5,139.24) ;
\draw [shift={(257.5,139.24)}, rotate = 180] [fill={rgb, 255:red, 0; green, 0; blue, 0 }  ][line width=0.08]  [draw opacity=0] (8.93,-4.29) -- (0,0) -- (8.93,4.29) -- cycle    ;
%Straight Lines [id:da9427188356855761]
\draw    (410.41,139.24) -- (434,139.24) ;
\draw [shift={(437,139.24)}, rotate = 180] [fill={rgb, 255:red, 0; green, 0; blue, 0 }  ][line width=0.08]  [draw opacity=0] (8.93,-4.29) -- (0,0) -- (8.93,4.29) -- cycle    ;
%Straight Lines [id:da5725605139132552]
\draw    (333.5,139.24) -- (357,139.24) ;
\draw [shift={(360,139.24)}, rotate = 180] [fill={rgb, 255:red, 0; green, 0; blue, 0 }  ][line width=0.08]  [draw opacity=0] (8.93,-4.29) -- (0,0) -- (8.93,4.29) -- cycle    ;
%Straight Lines [id:da98674425536353]
\draw    (54.5,152) -- (65.83,152.08) -- (77,152.02) ;
\draw [shift={(80,152)}, rotate = 539.69] [fill={rgb, 255:red, 0; green, 0; blue, 0 }  ][line width=0.08]  [draw opacity=0] (8.93,-4.29) -- (0,0) -- (8.93,4.29) -- cycle    ;
%Straight Lines [id:da1180143034702631]
\draw    (155.39,122.65) -- (166.63,122.65) -- (178.3,122.65) ;
\draw [shift={(181.3,122.65)}, rotate = 180] [fill={rgb, 255:red, 0; green, 0; blue, 0 }  ][line width=0.08]  [draw opacity=0] (8.93,-4.29) -- (0,0) -- (8.93,4.29) -- cycle    ;
%Straight Lines [id:da38991430633774415]
\draw    (155.39,151.82) -- (166.63,151.82) -- (178.3,151.82) ;
\draw [shift={(181.3,151.82)}, rotate = 180] [fill={rgb, 255:red, 0; green, 0; blue, 0 }  ][line width=0.08]  [draw opacity=0] (8.93,-4.29) -- (0,0) -- (8.93,4.29) -- cycle    ;
%Straight Lines [id:da47045088250050726]
\draw    (155.4,67.34) -- (166.88,67.34) -- (178.3,67.34) ;
\draw [shift={(181.3,67.34)}, rotate = 180] [fill={rgb, 255:red, 0; green, 0; blue, 0 }  ][line width=0.08]  [draw opacity=0] (8.93,-4.29) -- (0,0) -- (8.93,4.29) -- cycle    ;

% Text Node
\draw    (31.18,21.42) -- (54.18,21.42) -- (54.18,170.42) -- (31.18,170.42) -- cycle  ;
\draw (42.68,95.92) node  [font=\footnotesize,rotate=-270] [align=left] {\begin{minipage}[lt]{98.85pt}\setlength\topsep{0pt}
\begin{center}
\Gls{tmiv} encoder
\end{center}

\end{minipage}};
% Text Node
\draw    (80.4,26.18) -- (155.4,26.18) -- (155.4,48.18) -- (80.4,48.18) -- cycle  ;
\draw (117.9,37.18) node  [font=\footnotesize] [align=left] {\begin{minipage}[lt]{48.28pt}\setlength\topsep{0pt}
\begin{center}
\Gls{vtm} encoder
\end{center}

\end{minipage}};
% Text Node
\draw    (80.36,55.35) -- (155.36,55.35) -- (155.36,77.35) -- (80.36,77.35) -- cycle  ;
\draw (117.86,66.35) node  [font=\footnotesize] [align=left] {\begin{minipage}[lt]{48.28pt}\setlength\topsep{0pt}
\begin{center}
\Gls{vtm} encoder
\end{center}

\end{minipage}};
% Text Node
\draw (120,85.55) node [anchor=north west][inner sep=0.75pt]  [rotate=-90] [align=left] {\begin{minipage}[lt]{14.96pt}\setlength\topsep{0pt}
\begin{center}
{\Large ...}
\end{center}

\end{minipage}};
% Text Node
\draw    (181.62,21.37) -- (231.62,21.37) -- (231.62,85.37) -- (181.62,85.37) -- cycle  ;
\draw (206.62,53.37) node  [font=\footnotesize,rotate=-270] [align=left] {\begin{minipage}[lt]{40.79pt}\setlength\topsep{0pt}
\begin{center}
Subpicture\\bitstream\\merger
\end{center}

\end{minipage}};
% Text Node
\draw    (257.79,43.03) -- (333.79,43.03) -- (333.79,65.03) -- (257.79,65.03) -- cycle  ;
\draw (295.79,54.03) node  [font=\footnotesize] [align=left] {\begin{minipage}[lt]{48.62pt}\setlength\topsep{0pt}
\begin{center}
\Gls{vtm} decoder
\end{center}

\end{minipage}};
% Text Node
\draw (210,85.55) node [anchor=north west][inner sep=0.75pt]  [rotate=-90] [align=left] {\begin{minipage}[lt]{14.96pt}\setlength\topsep{0pt}
\begin{center}
{\Large ...}
\end{center}

\end{minipage}};
% Text Node
\draw (300,85.68) node [anchor=north west][inner sep=0.75pt]  [rotate=-90] [align=left] {\begin{minipage}[lt]{14.96pt}\setlength\topsep{0pt}
\begin{center}
{\Large ...}
\end{center}

\end{minipage}};
% Text Node
\draw    (437.31,21.85) -- (460.31,21.85) -- (460.31,171.85) -- (437.31,171.85) -- cycle  ;
\draw (448.81,96.85) node  [font=\footnotesize,rotate=-270] [align=left] {\begin{minipage}[lt]{99.53pt}\setlength\topsep{0pt}
\begin{center}
\Gls{tmiv} decoder
\end{center}

\end{minipage}};
% Text Node
\draw (390,85.94) node [anchor=north west][inner sep=0.75pt]  [rotate=-90] [align=left] {\begin{minipage}[lt]{14.96pt}\setlength\topsep{0pt}
\begin{center}
{\Large ...}
\end{center}

\end{minipage}};
% Text Node
\draw    (360.37,22.02) -- (410.37,22.02) -- (410.37,86.02) -- (360.37,86.02) -- cycle  ;
\draw (385.37,54.02) node  [font=\footnotesize,rotate=-270] [align=left] {\begin{minipage}[lt]{40.79pt}\setlength\topsep{0pt}
\begin{center}
Subpicture\\bitstream\\splitter
\end{center}

\end{minipage}};
% Text Node
\draw    (80.4,111.74) -- (155.4,111.74) -- (155.4,133.74) -- (80.4,133.74) -- cycle  ;
\draw (117.9,122.74) node  [font=\footnotesize] [align=left] {\begin{minipage}[lt]{48.28pt}\setlength\topsep{0pt}
\begin{center}
\Gls{vtm} encoder
\end{center}

\end{minipage}};
% Text Node
\draw    (80.36,140.91) -- (155.36,140.91) -- (155.36,162.91) -- (80.36,162.91) -- cycle  ;
\draw (117.86,151.91) node  [font=\footnotesize] [align=left] {\begin{minipage}[lt]{48.28pt}\setlength\topsep{0pt}
\begin{center}
\Gls{vtm} encoder
\end{center}

\end{minipage}};
% Text Node
\draw    (181.62,106.93) -- (231.62,106.93) -- (231.62,170.93) -- (181.62,170.93) -- cycle  ;
\draw (206.62,138.93) node  [font=\footnotesize,rotate=-270] [align=left] {\begin{minipage}[lt]{40.79pt}\setlength\topsep{0pt}
\begin{center}
Subpicture\\bitstream\\merger
\end{center}

\end{minipage}};
% Text Node
\draw    (257.79,128.59) -- (333.79,128.59) -- (333.79,150.59) -- (257.79,150.59) -- cycle  ;
\draw (295.79,139.59) node  [font=\footnotesize] [align=left] {\begin{minipage}[lt]{48.62pt}\setlength\topsep{0pt}
\begin{center}
\Gls{vtm} decoder
\end{center}

\end{minipage}};
% Text Node
\draw    (360.37,107.58) -- (410.37,107.58) -- (410.37,171.58) -- (360.37,171.58) -- cycle  ;
\draw (385.37,139.58) node  [font=\footnotesize,rotate=-270] [align=left] {\begin{minipage}[lt]{40.79pt}\setlength\topsep{0pt}
\begin{center}
Subpicture\\bitstream\\splitter
\end{center}

\end{minipage}};

\end{tikzpicture}}
  \caption{Simplified coding scheme using \gls{miv} and \gls{vvc} subpictures.}
  \label{fig:approach}
\end{figure*}
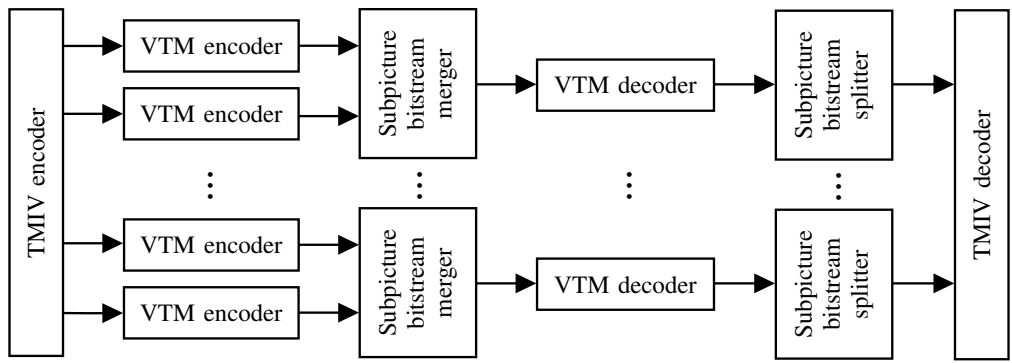

The encoding and decoding scheme is illustrated in Fig.~\ref{fig:approach}. The
encoding part consists of three steps. First, the texture and geometry atlases
are generated using the \gls{tmiv} encoder~\cite{cit:Salahieh2020}. Second, each
atlas is then compressed using the \gls{vtm}~\cite{cit:Chen2021}. Finally, each
pair of texture and geometry atlas is merged into a single bitstream using the
subpicture merge tool~\cite{cit:Hallapuro2020}, which is a part of the \gls{vtm}
package.

Rendering a viewport from the coded \gls{miv} bitstream involves first decoding
the packed atlases, extracting only the texture and geometry subpictures by
splitting the packed decoded atlases, and finally rendering the viewport using
the \gls{tmiv} decoder.

\subsection{Encoding}
The texture and geometry atlases are generated using
\gls{tmiv}~\cite{cit:Salahieh2020}. The size of a subpicture must be an integer
multiple of min \gls{cu} size, as specified by the \gls{vvc} standard.
Therefore, before \gls{vtm} encoding, each atlas is padded along columns and
rows to satisfy this constraint. Any rectangular empty space that remains after
packing of texture and geometry atlases is completed by generating a filler
sequence and encoding it to create a filler bitstream. The filler sequence
consists of YCbCr images with some constant value in the Y, Cb, and Cr planes.
The exact value in the filler image does not matter because it is neither
decoded nor used for rendering.

\subsection{Subpicture merging}
\Gls{vtm}~\cite{cit:Chen2021} encoder includes sample configurations that
enable the usage of the subpicture tool. It is possible to merge multiple
independent bitstreams into one, where each picture (frame) within the separate
bitstreams becomes a subpicture in the merged bitstream.

In the merging step, the coded texture, geometry and filler bitstreams are
merged using a subpicture bitstream merging tool provided by \gls{vtm}. A frame
of the resulting packed atlas, after merging, is shown in
Fig.~\ref{fig:packed-atlases}.

\subsection{Bitstream formatting}
After merging the three bitstreams, an \gls{miv} is presented with a \gls{v3c}
packed video component which is multiplexed to \gls{miv} bitstream. In order to
allow a client to correctly interpret the decoded data, an additional signalling
is included in the \gls{vps}. The extra signalling describes the type of each
region, its position and orientation in the packed frame. Additionally, a packed
independent \gls{sei} message can be included in the \gls{miv} bitstream. The
\gls{sei} message provides the mapping of regions in the packed video atlas to
\gls{vvc} subpicture identifiers, and it facilitates partial access of data in
the \gls{miv} bitstream.

\subsection{Decoding}
Arbitrary views from within the viewing volume can be generated by doing the
following: (1) decode the merged bitstream using \gls{vtm} decoder; (2) extract
the texture and geometry atlases from the decoded packed atlas using
information from the \gls{vps}; and (3) decode the remaining parts of the
\gls{miv} bitstream (i.e. \gls{v3c} atlas components) using the \gls{tmiv}
decoder and render the required viewport.

\section{Experimental evaluation}
\label{sec:experiments}
\begin{figure}[t]
  \centering
  \subfloat[ClassroomVideo]{\includegraphics[width=.225\textwidth]{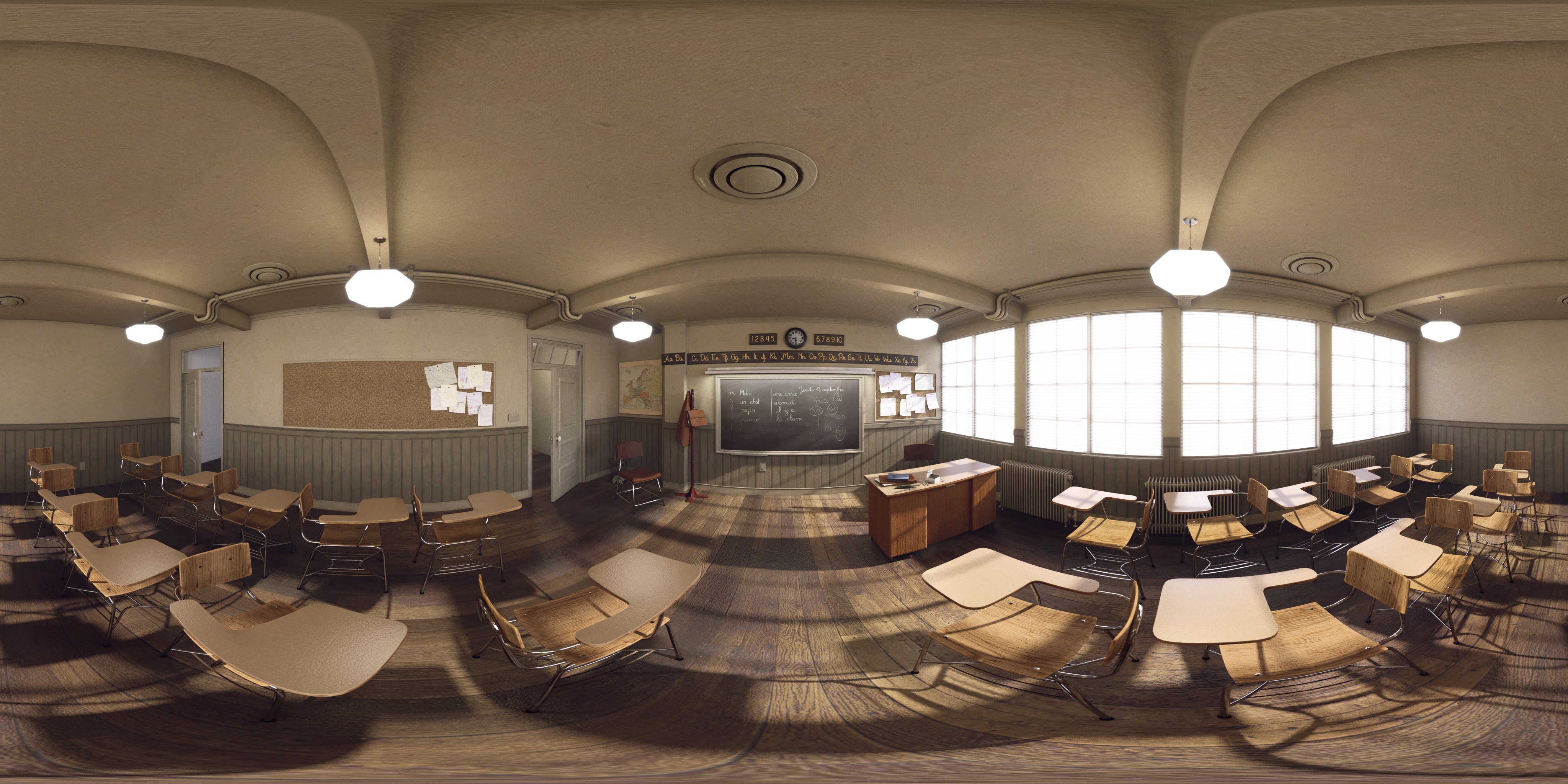}}
  \hspace*{.5cm}
  \subfloat[Frog]{\includegraphics[width=.2\textwidth]{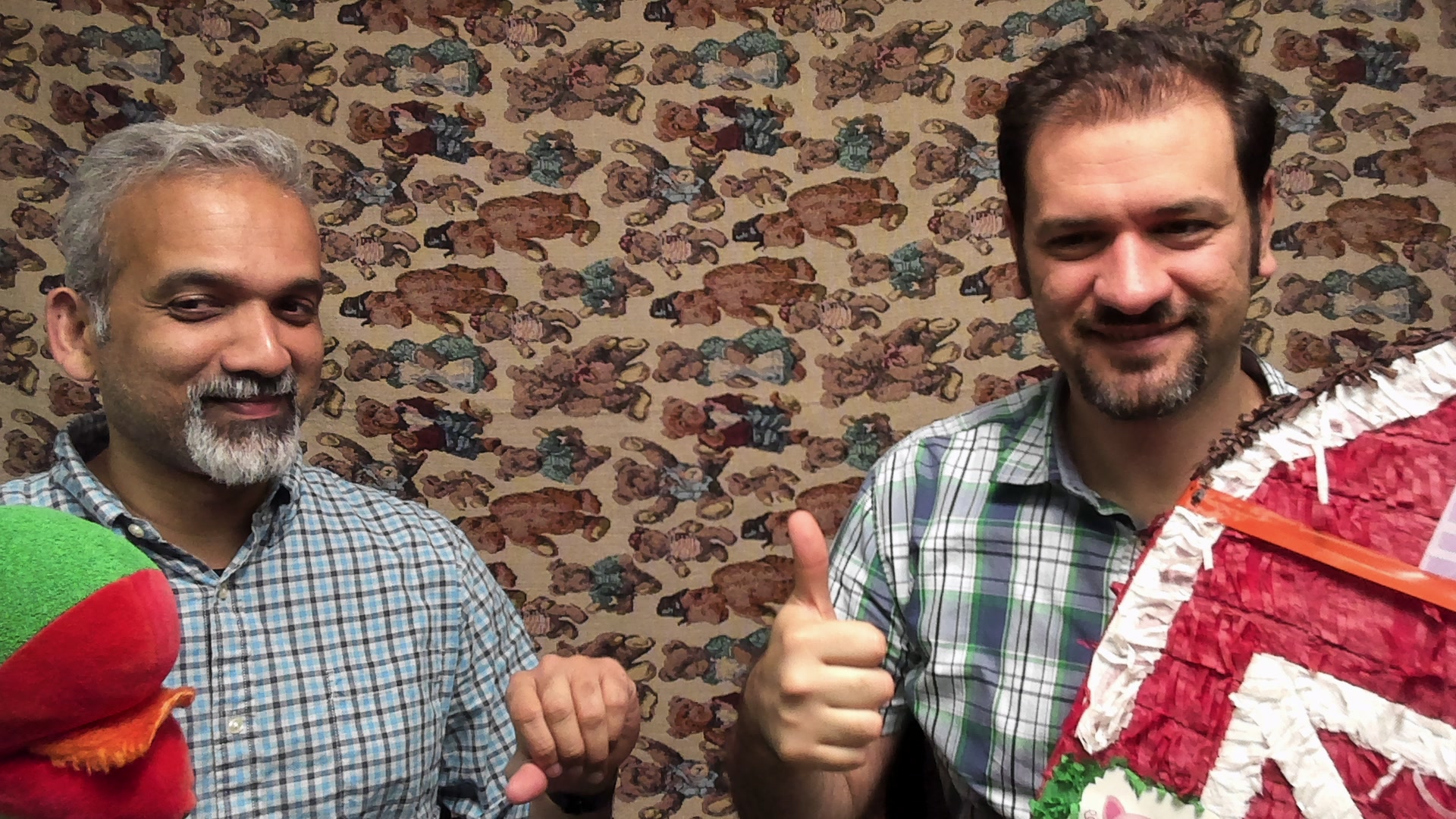}}
  \vspace*{.25cm}
  \subfloat[Chess]{\includegraphics[width=.15\textwidth]{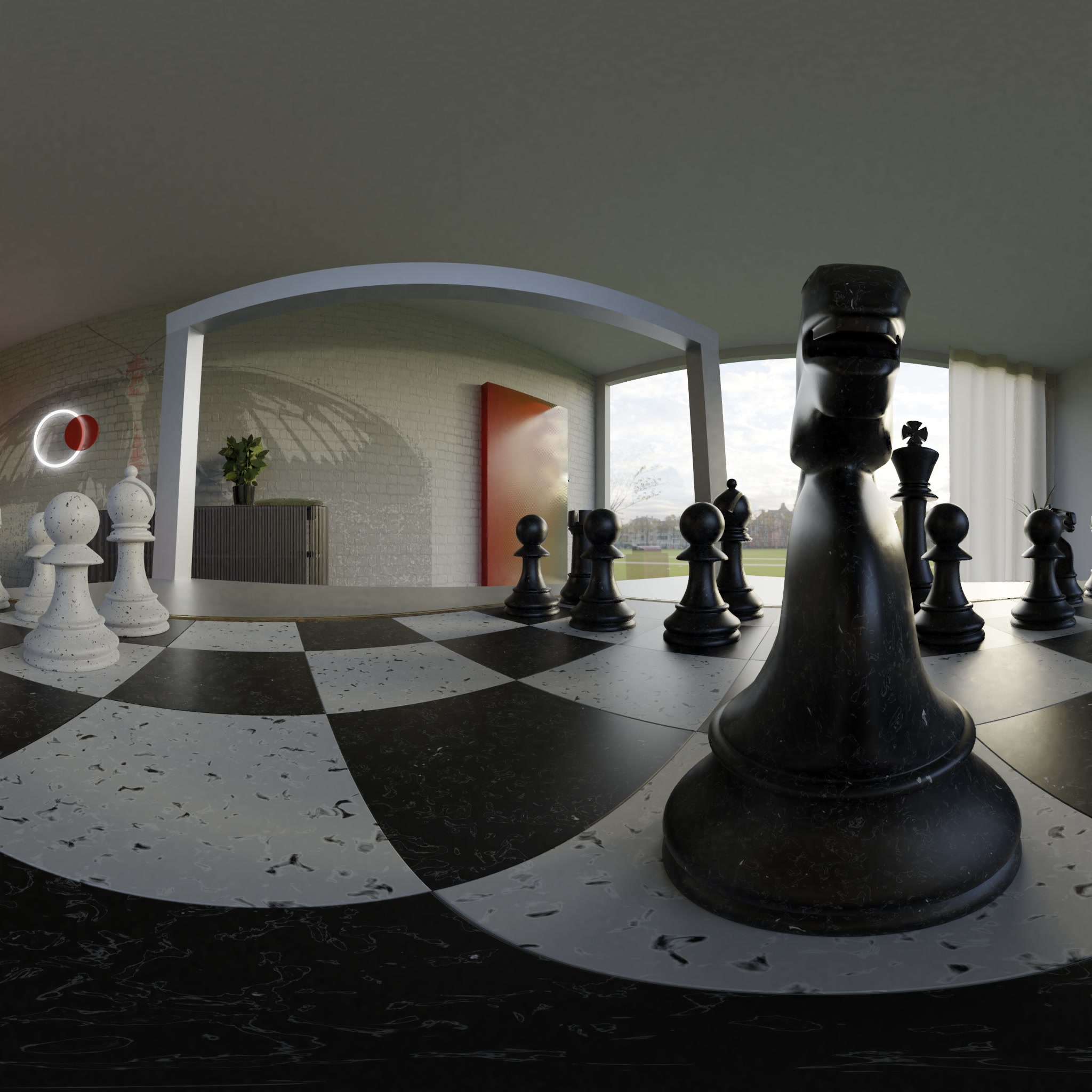}}
  \caption{One frame from one source camera of the sequences used.}
  \label{fig:sequences}
\end{figure}

MIV common test conditions document [3] defines evaluation methods as well as
test content. For the purpose of this paper, only a subset of the sequences was
evaluated; namely ClassroomVideo, Frog, and Chess, which are shown in
Fig.~\ref{fig:sequences} and their characteristics summarised in
Table~\ref{tab:sequences}.

\begin{table}[t]
  \centering
  \caption{Test sequences. E: \gls{erp}, P: perspective, \gls{cg}, \gls{nc}.}
  \label{tab:sequences}
  \begin{tabular}{l l l c c c}
    \toprule
    Sequence & \multicolumn{2}{c}{Type} & Resolution & Frame rate & Views \\
    \midrule
    ClassroomVideo & E & \gls{cg} & \( 4096 \times 2048 \) & \( 30 \) & \( 15 \) \\
    Frog & P & \gls{nc} & \( 1920 \times 1080 \) & \( 30 \) & \( 13 \) \\
    Chess & E & \gls{cg} & \( 2048 \times 2048 \) & \( 30 \) & \( 10 \) \\
    \bottomrule
  \end{tabular}
\end{table}

\glsreset{cg}
\glsreset{nc}

The \gls{cg} sequences contain near-perfect depth maps, whereas \gls{nc}
sequences contain estimated depth maps. Some sequences consist of \gls{erp}
views and others were captured with perspective cameras. Moreover, the
resolution varies from full \gls{hd} to \gls{uhd}.

The experiments were carried out under A17 configuration, meaning only \( 17 \)
frames were used. The atlases are generated with \gls{tmiv} 7.0 and are encoded
with \gls{vtm} 11.0 using random-access configuration along the parameters
displayed in Table~\ref{tab:vtm-params}. Each atlas is encoded using two sets
of \glspl{qp}, \gls{qp}1 to \gls{qp}4 (high bitrate) and \gls{qp}2 to \gls{qp}5
(low bitrate). Moreover, all experiments ran on cores Intel Xeon E5-2695 v2
@ 2.40GHz.

\begin{table}[t]
  \centering
  \caption{\gls{vtm} parameters used to encode atlases.}
  \label{tab:vtm-params}
  \begin{tabular}{p{6cm} c}
    \toprule
    Parameters & Value \\
    \midrule
    ALF; CCALF; LMCSEnable; JointCbCr; AMaxBT; EnablePartitionConstraintsOverride & \( 0 \) \\
    IBC; HashME; BDPCM & \( 1 \) \\
    \bottomrule
  \end{tabular}
\end{table}

The texture \glspl{qp} are sequence dependent, and they target \( 5 \) to
\( 50 \) \gls{mbps} bitrate range as defined in Table~\ref{tab:qps}. Each
texture \gls{qp} \( q \) is paired with a geometry \gls{qp} \( q' \) that is
calculated, as specified in the \gls{ctc}, as follows:

\begin{equation}
  q' = round( \max ( 1, -14.2 + 0.8 \cdot q ) )
\end{equation}

\begin{table}[t]
  \centering
  \caption{\gls{qp} used to encode texture atlases.}
  \label{tab:qps}
  \begin{tabular}{l l r r r r r}
    \toprule
    \multicolumn{2}{c}{Target bitrate [\gls{mbps}]} & \( 50 \) & \( 28 \) & \( 16 \) & \( 9 \) & \( 5 \) \\
    \midrule
    \multicolumn{2}{c}{\gls{qp} index} & \( 1 \) & \( 2 \) & \( 3 \) & \( 4 \) & \( 5 \) \\
    \midrule
    \multirowcell{3}{Sequence} & ClassroomVideo & \( 25 \) & \( 27 \) & \( 30 \) & \( 33 \) & \( 38 \) \\
    & Frog & \( 30 \) & \( 36 \) & \( 43 \) & \( 47 \) & \( 51 \) \\
    & Chess & \( 11 \) & \( 18 \) & \( 25 \) & \( 31 \) & \( 38 \) \\
    \bottomrule
  \end{tabular}
\end{table}

The quality of the reconstructed source-views is evaluated by computing
objective quality metrics. Specifically, the metrics used are the \gls{psnr},
the \gls{ws-psnr}~\cite{cit:Sun2017}, the \gls{iv-psnr}~\cite{cit:Dziembowski2019},
and the \gls{vmaf}~\cite{cit:Li2016}. Furthermore, the
\gls{bd-rate}~\cite{cit:Bjontegaard2001} is computed for each of these metrics.

\begin{table*}[t!]
  \centering
  \caption{\gls{bd-rate} over \gls{tmiv} 7.0 + \gls{vtm} 11.0.}
  \label{tab:bd-rate}
  \begin{tabular}{l c c c c c c c}
    \toprule
    Sequence &
    \makecell{High bitrate\\ \gls{bd-rate} Y\\ \gls{ws-psnr}} &
    \makecell{Low bitrate\\ \gls{bd-rate} Y\\ \gls{ws-psnr}} &
    \makecell{High bitrate\\ \gls{bd-rate}\\ \gls{vmaf}} &
    \makecell{Low bitrate\\ \gls{bd-rate}\\ \gls{vmaf}} &
    \makecell{High bitrate\\ \gls{bd-rate}\\ \gls{iv-psnr}} &
    \makecell{Low bitrate\\ \gls{bd-rate}\\ \gls{iv-psnr}} \\
    \midrule
    ClassroomVideo & \( 0.2\% \) & \( 0.4\% \) & \( 0.1\% \) & \( 0.4\% \) & \( 0.1\% \) & \( 0.3\% \) \\
    Frog & \( 0.4\% \) & \( 0.4\% \) & \( 0.2\% \) & \( 0.3\% \) & \( 0.3\% \) & \( 0.4\% \) \\
    Chess & \( 0.1\% \) & \( 0.2\% \) & \( 0.1\% \) & \( 0.3\% \) & \( 0.1\% \) & \( 0.2\% \) \\
    \bottomrule
  \end{tabular}
\end{table*}

\begin{table*}[t]
  \centering
  \caption{Runtime ratios.}
  \label{tab:runtime}
  \begin{tabular}{l r r r r r r r r}
    \toprule
    Sequence & \makecell{\gls{tmiv}\\ encoding}
    & \makecell{\gls{tmiv}\\ encoding\\ \& pad}
    & \makecell{\gls{vtm}\\ encoding}
    & \makecell{\gls{vtm}\\ encoding\\ \& merge}
    & \makecell{\gls{vtm}\\ decoding}
    & \makecell{\gls{vtm}\\ decoding\\ \& split}
    & \makecell{\gls{tmiv}\\ decoding} \\
    \midrule
    ClassroomVideo & \( 99.6\% \) & \( 99.6\% \) & \( 99.5\% \) & \( 99.5\% \) & \( 190.6\% \) & \( 512.3\% \) & \( 97.4 \% \) \\
    Frog & \( 98.3\% \) & \( 98.3\% \) & \( 98.3\% \) & \( 98.3\% \) & \( 274.0\% \) & \( 396.0\% \) & \( 98.2\% \) \\
    Chess & \( 98.0\% \) & \( 98.0\% \) & \( 97.9\% \) & \( 97.9\% \) & \( 272.5\% \) & \( 635.9\% \) & \( 97.2\% \) \\
    \bottomrule
  \end{tabular}
\end{table*}

The \gls{bd-rate} results are summarised in Table~\ref{tab:bd-rate}.
The proposed approach incurs in negligible coding losses, up to \( 0.4\% \),
due to the overhead associated to the filler video coded data required to create
the merged bitstream (\gls{irap} subpictures use in average \( 236 \) bytes and
inter-coded subpictures use in average \( 14 \) bytes). The results are slightly
higher for Frog as the overhead includes the filler video and the texture
and geometry atlases are padded along the width and height. Nevertheless, the
quality of reconstructed source-views is almost identical between the anchor
and the proposed approach.

The complexity of the approach is based on the runtime ratios
(Table~\ref{tab:runtime}). There are three main outcomes. First, the \gls{tmiv}
encoding and the \gls{tmiv} decoding have similar complexities as these two
operations are changeless. Second, the complexity of the padding operation,
encoding the filler with \gls{vtm} and the merging operation are almost
negligible. Finally, the more demanding operations are the decoding of the
merged bitstreams with \gls{vtm} and extracting afterwards the texture and
geometry atlases. Note that these two last operations can be improved by
extracting and decoding only the required subpictures.

The overall pixel rate of the merged bitstream can be reduced by changing the
orientation of the atlases, in order to reduce the dimensions of the filler
component.

Since it is possible to maintain the quality of reconstructed source-views,
coding texture and geometry atlases as subpictures within the same bitstream
is a feasible approach.

\section{Conclusion}
\label{sec:conclusion}
This paper used the subpicture tool of \gls{vvc} to code \gls{6dof} immersive
video. The independently extractable and decodable property of subpictures
provides additional flexibility in the coding of \gls{6dof} immersive videos,
some of which were elaborated upon.

\balance

A specific case, where the texture and geometry atlases generated by \gls{miv}
are coded as subpictures within the same atlas, was selected for experimentation.
The results showed that while there is a small (\( 0.1\% \) to \( 0.4\% \))
\gls{bd-rate} overhead incurred, the number of decoder instantiations required
is reduced by half. The evaluated approach also provides the ability to
independently encode, extract, and decode parts of the coded bitstream,
thus improving parallelism.

While, this paper focused on one specific case, it is evident that there may
be many other cases, specifically for coding large dynamic \gls{6dof}
volumetric videos, where subpictures could be found useful.

\bibliographystyle{IEEEtran}
\bibliography{references}

\end{document}